\documentclass[singlecolumn,epjc3]{svjour3}
\smartqed  
\RequirePackage{graphicx}
\journalname{Eur. Phys. J. C}

\begin{document}
\title{Anisotropic fluid spheres of embedding class one using Karmarkar condition}
	
	\author{S.K. Maurya\thanksref{e1,addr1}
     \and S. D. Maharaj \thanksref{e2,addr2}}

	\thankstext{e1}{e-mail: sunil@unizwa.edu.om}
    \thankstext{e2}{e-mail: maharaj@ukzn.ac.za}

	\institute{Department of Mathematical \& Physical Sciences, College of Arts \& Science,
		University of Nizwa, Nizwa, Sultanate of Oman\label{addr1} \and Astrophysics and Cosmology Research Unit, School
of Mathematics, Statistics and Computer Science, University
of KwaZulu-Natal, Private Bag X54001, Durban 4000,
South Africa\label{addr2}}

	\date{Received: date / Accepted: date}
	
	\maketitle	
\begin{abstract}

We obtain a new anisotropic solution for spherically symmetric spacetimes
 by analysing of the Karmarkar embedding condition. For this purpose we construct a suitable form of one of the
 gravitational potentials to obtain a closed form solution. This form of the remaining gravitational potential
allows us to solve the embedding equation and integrate the field equations. The resulting new anisotropic solution is well behaved which can be utilized to construct realistic static fluid spheres.
Also we estimated masses and radii of fluid spheres for LMC X-4 and EXO 1785-248 by using observational data sets values. The obtained masses and radii show that our anisotropic solution can represent fluid spheres to a very good degree of accuracy.

	\end{abstract}


\section{Introduction}

The presence of nonzero anisotropy, in which the radial pressure differs from the tangential pressure,  is an important component  in modeling relativistic stellar systems in the absence of the electric field. The pioneering paper by Bowers and Liang \cite{bowers} introduced anisotropic spheres in general relativity. Subsequently there has been extensive research in studying the physics related to anisotropic pressures. It was shown by Dev and Gleiser \cite{dev1,dev2} that pressure anisotropy influences the mass, structure and physical properties of highly compact  spheres. It is important to observe that the mass of the object and the redshift both change with different values of the magnitude of  the anisotropy. In recent times there has been considerable effort in modeling observed astrophysical objects in the presence of anisotropy. Some recent research papers addressing this physical feature include the works of Sharma and Ratanpal \cite{sharma}, Ngubelanga et al \cite{ngubelanga1,ngubelanga2}, Sunzu et al \cite{sunzu1,sunzu2}, Murad and Fatema \cite{murad1,murad2} and Murad \cite{murad3}, and references contained therein. The physical analyses contained in these treatments confirm the importance of including nonzero anisotropy in modeling astrophysical objects. It should be noted that the presence of anisotropic pressures leads to values of observed compactness parameters for several astrophysical bodies including Her X-1, PSR 0943+10, 4U 1820-30, SAX J 1808.4-3658, and 4U 1728-34. It is therefore important to generate analytical models of the Einstein field equations, with a matter tensor containing anisotropy, which are consistent with physical requirements for astrophysical applications.

To generate a physically consistent model we need to find an analytical solution of the general relativistic field equations.  This is achieved by restricting the spacetime geometry, the matter content or specifying an equation of state. A rather different  approach is to use the embedding of a four-dimensional differentiable manifold into a higher dimensional Euclidean space. Embedding of curved spacetimes into spacetimes of higher flat dimensions has assisted in generating several new exact models  in cosmology and relativistic astrophysics \cite{stephani}. The embedding condition gives an additional differential equation, the so-called Karmarkar condition \cite{karmarkar},  in static spherical spacetimes relating the two gravitational potentials. A solution of the Karmarkar condition then helps to solve the Einstein field equations. This method has proved to be a fruitful mechanism to find new relativistic astrophysical models in recent investigations. Bhar et al \cite{bhar}, Maurya et al  \cite{maurya1,maurya2,maurya3,maurya4}, Newtonsingh et al \cite{newton1} and Newton Singh and Pant \cite{newton2} have generated different families of physically acceptable Karmarkar spacetimes that describe the interior regions of acceptable stars. In these analyses a particular form of one of the potentials is made which solves the Karmarkar condition, giving the second potential, eventually leading to an astrophysical model. In our treatment we show that a very general form of the chosen potential, including hyperbolic functions, leads to an astrophysical model with desirable physical features. This shows that the Karmarkar condition allows for more complicated (and acceptable) physical behaviour than the earlier simpler forms used for the gravitational potentials.

In this paper we present a new interior anisotropic model for astrophysical compact stars by solving the embedding condition in static spherical spacetimes. We show that the resulting exact solutions can be used to describe a physically reasonable astrophysical matter distribution. The exterior region is characterized by the Schwarzschild vacuum metric. We discuss the Einstein field equations in Sect. 2 and present the Karmakar embedding condition. In Sect. 3, we generate an exact solution to the embedding condition and show how this leads to an exact solution of the field equations. In Sect. 4 we present the matching conditions between interior and exterior spacetimes regions; we also demonstrate that the parameters arising are bounded. A detailed physical analysis is undertaken in Sect. 5. In particular the stability, cracking and energy conditions of the relativistic sphere are studied. We also investigate the physical features of the model in connection with the objects LMC X-4 and EXO 1785-248; the results are presented in the form of tables and graphs. We discuss the significance of the results obtained in this paper in Sect. 6.

\section{Field equations and the Karmarkar condition}

We assume that the interior matter of the star is locally anisotropic which is given by the following line element (by taking $c=1$)

\begin{equation}
ds^{2} =-e^{\lambda(r) } dr^{2} -r^{2} (d\theta ^{2} +\sin ^{2} \theta \, d\phi ^{2} )+e^{\nu(r) } \, dt^{2} ,
\label{1}
\end{equation}

\noindent where, $e^{\lambda(r)}$ and $e^{\nu(r)}$ represent the gravitational potentials of stellar structure. The Einstein field equations corresponding to an anisotropic fluid distribution is given by  (assume $G=c=1$)

\begin{equation}
-8\pi\,T^{i}_{j}= R^{i}_{j}-\frac{1}{2}\,R\,g^{i}_{j}, \label{field}
\end{equation}

\noindent where, $T^{i}_{j}$ and  $R^i_j$ represent energy momentum tensor and contracted Ricci tensor respectively while $R$ is the scalar curvature tensor. The energy tensor for the anisotropic matter distribution can be defined as

\begin{equation}
\label{3}
T^{i}_{j}=(\rho+p_t)\,v^i\,v_j-p_t\,g^i_j+(p_r-p_t)\,u^i\,u_j .
\end{equation}

\noindent where the contravariant quantity $v^i$ is the four-velocity vector and $u^i$ is the unit spacelike vector in the radial direction. Here $p_r$, $p_t$ and $\rho$ denote the radial pressure, tangential pressure and matter density for anisotropic matter.

In view of line element (1), the Einstein field equations (\ref{field}) provide the following differential equations for the anisotropic star as

\begin{eqnarray}
\rho & =&\frac{e^{-\lambda}}{8\pi}\left[\frac{r\,\lambda '+e^{\lambda}-1 }{r^{2}}\right], \label{6}\\
p_{r} &=&\frac{e^{-\lambda}}{8\pi}\left[\frac{r\,v'-e^{\lambda}+1}{r^{2}} \right] ,\label{4} \\
p_{t} &=&\frac{e^{-\lambda}}{8\pi}\left[\frac{2r\,v''-r\,\lambda'v'+r\,v'^{2} +2\,v'-2\,\lambda'}{4r^2} \right] .\label{5}
\end{eqnarray}

\noindent Here primes denote the derivative with respect to the radial coordinate $r$. The value of the velocity of light ($c$) and the gravitational constant ($G$) are taken to be unity in the above coupled differential equations. Furthermore we obtain the anisotropic factor by using the pressure isotropy condition with Eqs.(\ref{4}) and (\ref{5}) as

\begin{equation}
\label{7}
\Delta =\,p_{t} -\, p_{r} \,=\frac{e^{-\lambda }}{8\pi}\left[\frac{2r\,v''-r\,\lambda 'v'+ r\,v'^{2}-2v'-2\lambda '}{4r}\right] -\frac{e^{-\lambda }-1}{r^{2}}.
\end{equation}

\subsection{Karmarkar condition}
 It is well known that the spherical symmetric line element (1) can always be  embedded in six dimensional flat spacetime which implies that the spherical symmetric line element is of embedding class two in general.  On the other hand we can also embed the spherical line element into five dimensional flat spacetime if it satisfies the Karmarkar condition \cite{karmarkar}. Then it represents the spacetime of embedding class one. However it is a necessary and sufficient condition for the spherically symmetric spacetime to be a class one. The Karmarkar condition is given in terms of curvature components by

\begin{equation}
\Re_{1414}=\frac{\Re_{1212}\,\Re_{3434}+\Re_{1224}\,\Re_{1334}}{\Re_{2323}}, \label{karmarkar}
\end{equation}
with $\Re_{2323}\ne0$ \, \cite{Pandey1982}. The nonzero components of the Riemann curvature tensor $\Re_{hijk}$ for the metric (1) are given by\\

$\Re_{2323}=\frac{sin^2\theta\,(e^{\lambda}-1)\,r^2}{e^{\lambda}}$, \,\, $\Re_{1212}=\frac{\lambda'\,r}{2}$,\,\,  $\Re_{2424}=\frac{\nu'\,r\,e^{\nu-\lambda}}{2}$,\\

$\Re_{1224}=0$,\,\,$\Re_{1414}=\frac{e^\nu}{4}\,[2\,\nu''+\nu'^2-\lambda'\,\nu']$,\,\, \,$\Re_{3434}=sin^2{\theta}\,\Re_{2424}$.\\

By plugging these components of $\Re_{hijk}$ in Eq.(\ref{karmarkar}) we get following differential equation

\begin{equation}
\label{9}
\frac{\nu''}{\nu'}+\frac{\nu'}{2}=\frac{\lambda'\,e^{\lambda}}{2\,(e^{\lambda}-1)}.
\end{equation}
On solving the differential equation (\ref{9}) we obtain the potential

\begin{equation}
\label{10}
e^{\nu}=\left[C+D\int{\sqrt{(e^{\lambda(r)}-1)}dr}\right]^2,
\end{equation}
 where,  $C$ and $D$  are nonzero arbitrary constants of integration.

\subsection{Tolman-Oppenheimer-Volkoff equation}

Now to derive the Tolman-Oppenheimer-Volkoff (TOV) equation we evaluate $p_r+\rho$ from eqs.(\ref{6}) and (\ref{4}) to get

\begin{equation}
	 \rho+p_r= \frac{(\lambda'+\nu')\,e^{-\lambda}}{8\,\pi\,r}. \label{t1}
\end{equation}
The derivative of the radial pressure is

\begin{equation}	
\frac{dp_r}{dr}=\left[\frac{r\,\nu''-r\,\nu'\,\lambda'-\nu'-\lambda'}{8\,\pi\,r^2}\right]\,e^{-\lambda}+\frac{2(1-e^{-\lambda})}{8\,\pi\,r^3}. \label{t2}
\end{equation}
Then using Eqs.(\ref{7}),(\ref{t1}) and (\ref{t2}) we get

\begin{equation}
\frac{2}{r}(p_t-p_r)-\frac{dp_r}{dr}-\frac{1}{2}\,\nu'\,(\rho+p_r)=0. \label{TOV1}
\end{equation}

\noindent If the gravitational mass within a compact star of radius $r$ is denoted as $M_G(r)$, then it can be given by Tolman-Whittaker formula as (\cite{Landau})

\begin{equation}
M_G(r)=\frac{1}{2}r^2e^{\frac{\nu-\lambda}{2}\nu'}. \, \label{TOV2}
\end{equation}

Then from Eqs.(\ref{TOV1}) and (\ref{TOV2}), we obtain

\begin{equation}
\frac{2}{r}(p_t-p_r)=\frac{dp_r}{dr}+\frac{M_G(r)\,(\rho+p_r)}{r^2}\,e^{\lambda-\nu}. \label{TOV}
\end{equation}

\noindent The above equation (\ref{TOV}) represents the well known generalized Tolman-Oppenheimer-Volkoff (TOV) equation which provides the equilibrium condition for anisotropic stellar matter distribution.

\section{Anisotropic solution of embedding class one}

 Eqs.(\ref{6}-\ref{4}) have five unknowns namely $\nu$, $\lambda$, $\rho$, $p_r$ and $p_t$. However the Karmarkar condition provides a relation between $\nu$ and $\lambda$ which implies that we have four conditions (including three equations) to solve this system of equations. For this paper we consider a totally new expression for gravitational potential $e^{\lambda}$ which has not been used before. We take
\begin{equation}
e^{\lambda}=\frac{1+2cr^2+\cosh[2(ar^2+b)]}{1+\cosh[2(ar^2+b)]},  \label{lambda}
\end{equation}

\noindent where $a$, $b$ and $c$ are nonzero constants, The units of the constants $a$ and $c$ are $length^{-2}$. We need to check whether the given expression for $\lambda$ is physically valid. For this purpose we obtain $e^{\lambda}$ at the centre and plot  Fig. 1. We observe that  it is increasing monotonically away from centre and $e^{\lambda}=1$ at centre. This behavior of $e^{\lambda}$ indicates that it is physically acceptable.

By plugging the value of $\lambda$ into Eq.(\ref{10}), we obtain

\begin{equation}
e^{\nu}=A^2\left[B+\tan^{-1}\sinh(ar^2+b)\right]^2,  \label{nu}
\end{equation}
 where, $A=D\,\frac{\sqrt{c}}{2\,a}$ and $B=\frac{2\,a\,C}{\sqrt{c}\,D}.$
 The function $e^{\nu}=A^2\,[B+\tan^{-1}\sinh\,b]^2$ is finite and positive at centre. Also we may observe form Fig.1 that the gravitational potential $e^{\nu}$ is increasing with $r$ throughout the star. This implies that above the expression of $\nu$ may be suitable to obtain physically valid anisotropic solution according to Lake \cite{Lake2003}.

\begin{figure}[h]
\centering
\includegraphics[width=5.5cm]{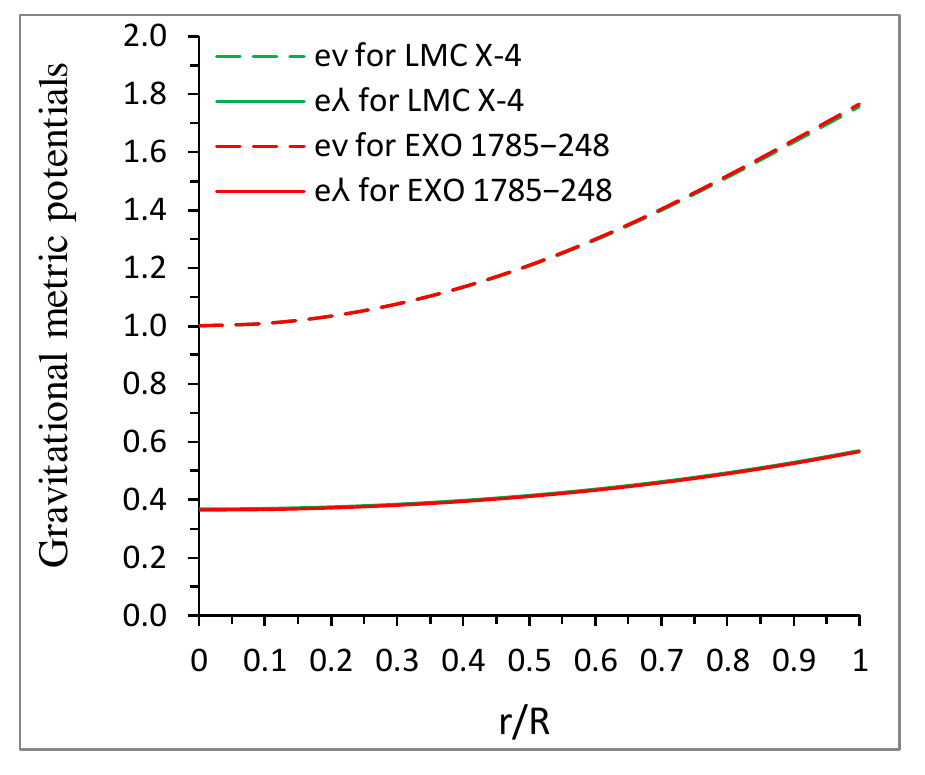}
\caption{Behavior of gravitational potential $e^{\nu}$ (solid line) and $e^{\lambda}$ (dotted line) verses fractional radius $r/R$ for LMC X-4 and EXO 1785-248. For plotting of this figure the numerical values of physical parameters and constants are as follows:(i) $a=0.004$, $b=0.0021$, $c=0.0107$, $A=0.4806$, $B=1.2607$, $M=1.29 M_{\odot}$, and $R=8.831 km$ for LMC X-4, (ii). $ a=0.00393$, $b= 0.0025$, $c=0.01074$, $A=0.4905$, $B=1.2293$,  $M=1.3 M_{\odot}$ and $R=8.849 km$ for EXO 1785-248. These numerical values are given in Table 1.}
\end{figure}

The expressions for the matter density, radial and tangential pressures are obtained (by taking $\psi=a\,r^2+b$, \,  $\Phi(r)=B+\tan^{-1}\sinh\psi$) as

\begin{equation}
\rho=\frac{2\,c\,(2+3\,c\,r^2+3\,\cosh2\psi-4\,a\,r^2\,\sinh2\psi)}{8\pi\,(1+2\,c\,r^2+\cosh2\psi)^2},
\end{equation}

\begin{equation}
p_r=\frac{2\,[-c\,B-c\,\tan^{-1}\sinh\psi+4a\,\cosh\psi]}{8\pi\,(B+\tan^{-1}\sinh\psi)\,(1+2\,c\,r^2+\cosh2\psi)},
\end{equation}

\begin{equation}
p_t=\frac{4\,\cosh\psi[2a\,(1+cr^2+\cosh2\psi+c\,r^2\,\Phi(r)\,\sinh\psi-a\,r^2\,\sinh2\psi)-c\,\Phi(r)\,\cosh\psi]}{8\pi\,(B+\tan^{-1}\sinh\psi)\,(1+2\,c\,r^2+\cosh2\psi)^2}.
\end{equation}

\begin{figure}[h]
\centering
\includegraphics[width=5.5cm]{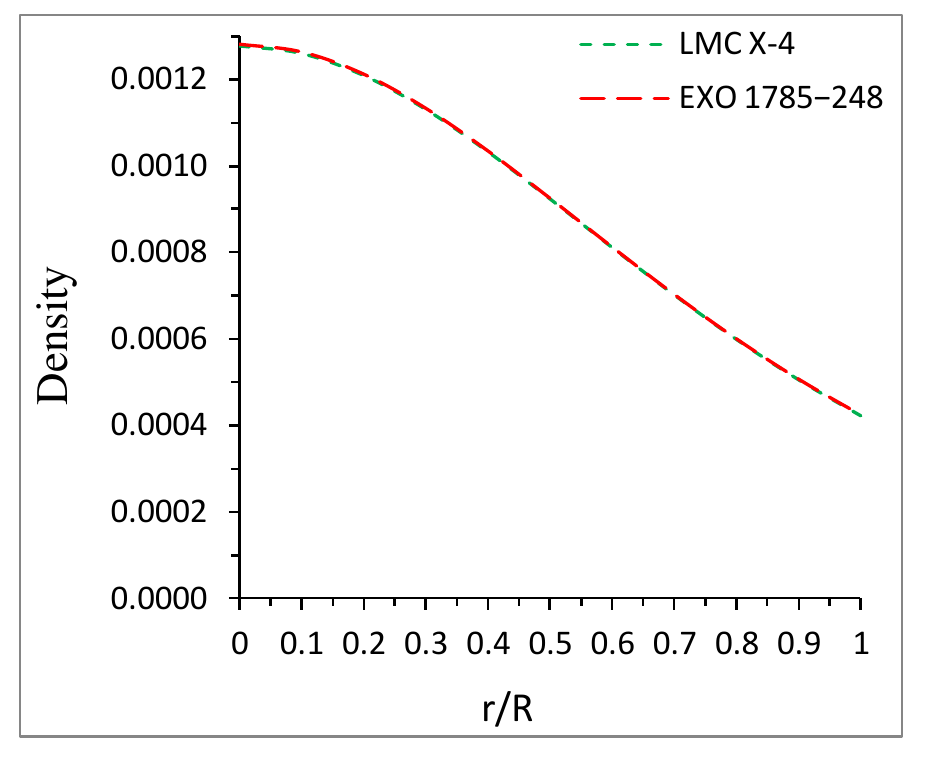} \includegraphics[width=5.5cm]{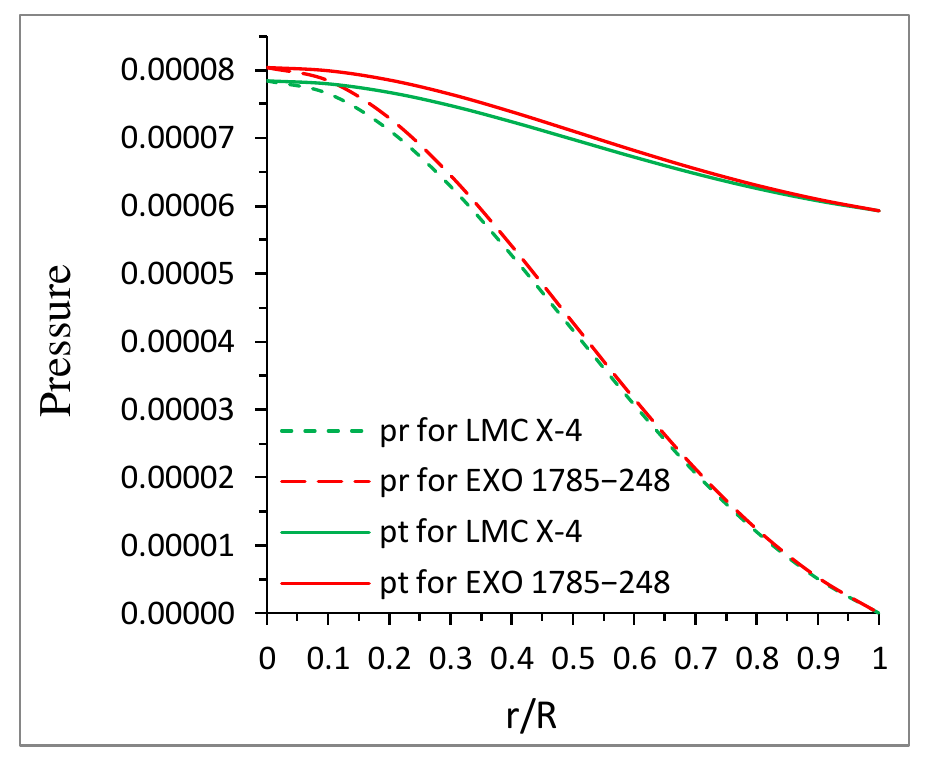}
\caption{Behavior of energy density $\rho$ (left panel), and radial pressure, $p_r$ dotted line, tangential pressure, $p_t$ solid line (right panel) verses fractional radius $r/R$ for LMC X-4 and EXO 1785-248. For plotting of this figure we have employed data set values of physical parameters and constants as used in Fig. 1.}
\end{figure}

We plot the variation of matter density, radial and tangential pressures in Fig.2. We can see the density is maximum at the centre and the minimum occurs on the boundary of the star. The radial and tangential pressures are both monotonically decreasing away from the centre. However $p_r$ become zero at the boundary of the star which gives the radius of the star. Note that $p_t$ is nonzero and positive. It is worth pointing out here that the central density has a order of  $10^{15} gm/cm^3$ which indicates that the nuclear matter is more appropriate for the anisotropic fluid distribution. The expression for anisotropic factor is given by

\begin{equation}
\Delta=\frac{4\,r^2\,(c\,\Phi(r)-2\,a\,\cosh\psi)(c+a\,\sinh2\psi)}{8\pi\,(B+\tan^{-1}\sinh\psi)\,(1+2\,c\,r^2+\cosh2\psi)^2}.
\end{equation}

\begin{figure}[h]
\centering
\includegraphics[width=5.5cm]{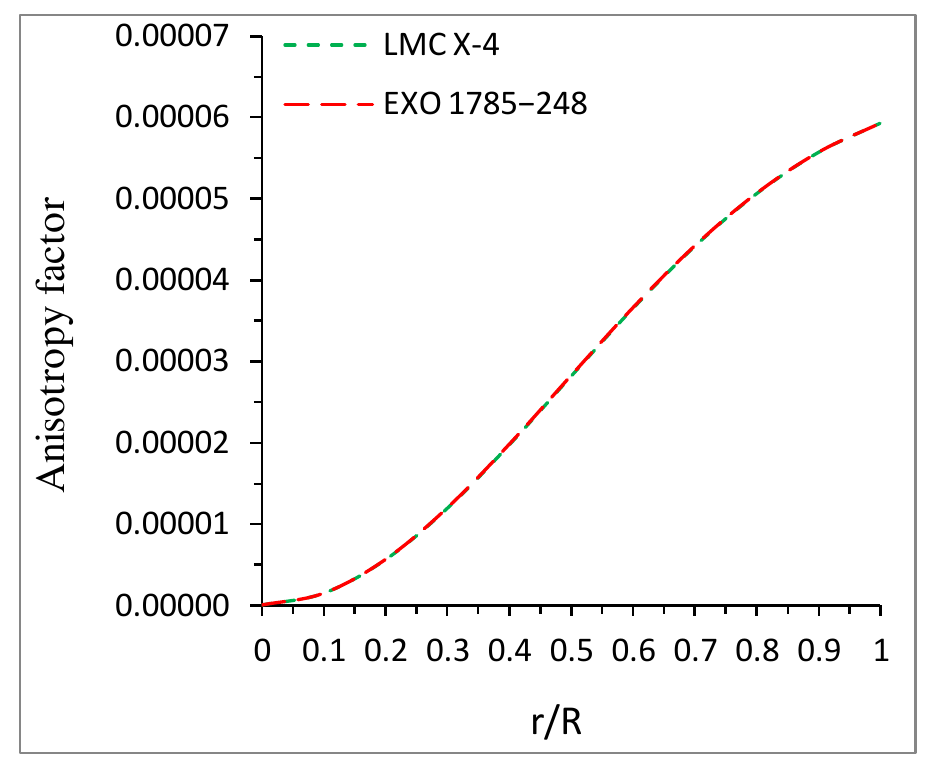}
\caption{Behavior of anisotropy factor $\Delta$  verses fractional radius $r/R$ for LMC X-4 and EXO 1785-248. For plotting of this figure we have employed data set values of physical parameters and constants as used in Fig.1 and 2.}
\end{figure}

\noindent The pressure anisotropy of the matter plays an important role in determining the stability of the model. Since the force due to anisotropy can be defined as $F=\frac{2\,\Delta}{r}$, which implies if $\Delta$ is positive then the direction of the force is outward; however the force will be directed inward if $\Delta$ is negative. But the existence of a repulsive force (in the case in which tangential pressure is more than radial pressure i.e. $p_t > p_r$) allows the construction of more compact star models when using anisotropic fluid than when using isotropic fluid \cite{Gokhroo}. From   Fig.3, we can see that $\Delta$ is positive and finite throughout inside the star. Also it is zero at the centre and attains a maximum at the boundary of star. \\

The gradients of pressure and density are given by

\begin{equation}
\frac{dp_r}{dr}=\frac{2\,r\,\left[\,p_{1}\,p_{2}\,(a+2\,p_{3}\,\Phi(r)\,\cosh\psi)+a\,p_{r2}\,p_{4}\,\Phi(r)\,\right]}
{8\pi\,a\,\cosh\psi\,(B+\tan^{-1}\sinh\psi)^2\,(1+2\,c\,r^2+\cosh2\psi)^2},
\end{equation}

\begin{equation}
\frac{dp_t}{dr}=\frac{2\,r\,\left[\,a\,p_{2}\,p_{5}-a\,p_{2}\,p_{5}\,\Phi(r)+4\,p_{3}\,p_{5}\,\Phi(r)\,\cosh\psi-a\,p_{2}\,(p_{6}+p_{7})\,\Phi(r)\,\right]}
{8\pi\,a\,(B+\tan^{-1}\sinh\psi)^2\,(1+2\,c\,r^2+\cosh2\psi)^3}
\end{equation}

\begin{equation}
\frac{d\rho}{dr}=\frac{4\,c\,r\,\left[\,2\,\rho_{1}\,(\,c\,\Phi(r)-4\,a\,\cosh\psi\,)-4\,\rho_{2}\,(c+a\,\sinh2\psi)\right]}
{8\pi\,(1+2\,c\,r^2+\cosh2\psi)^3}.
\end{equation}
In the above we have set\\

$p_{1}=[\,c\,\Phi(r)-4\,a\,\cosh\psi\,]$, \,\, $p_{2}=[\,1+2\,c\,r^2+\cosh2\psi\,]$,\,\\

\,$p_{3}=[\,c+a\,\sinh2\psi\,]$,\,\,\, $p_{4}=[\,-c+2\,a\,\sinh2\psi\,]$, \,\, \\

$p_5=[\,c\,\Phi(r)\,\cosh\psi-2a\,(1+c\,r^2+\cosh2\psi+c\,r^2\,\Phi(r)\,\sinh\psi-a\,r^2\,\sinh2\psi)\,]$,\\

$p_{6}=\left[\,a\,c\,r^2\,\Phi(r)+(c-2\,a^2\,r^2)\,\cosh\psi+a\,c\,r^2\,\Phi(r)\,\cosh2\psi\,+2\,a^2\,r^2\,\cosh3\psi\right]$, \\

$p_{7}=\left[\,-a\,\sinh\psi-a\,\sinh\psi-2\,a\,c\,r^2\,\sinh\psi-c\,\Phi(r)\,\sinh\psi\,\cosh\psi-a\,\sinh3\psi\,\right]$,\\

$\rho_1=[\,c-4a^2\,r^2\,\cosh2\psi+a\,\sinh2\psi\,]$, \,\, $\rho_2=[\,3+2\,c\,r^2+3\,\cosh2\psi-4\,a\,r^2\,\sinh2\psi\,]$.

\section{Bounds on the parameters and matching conditions}
\subsection{Bounds on the parameters}

Since radial pressure $p_r$ and the tangential pressure $p_t$ are positive and finite inside the star we obtain the upper bound of $B$ as

\begin{equation}
B\,\,< \,\left(\,\frac{4\,a\,\cosh (b)}{c}-\tan^{-1}\sinh(b)\,\right).  \label{B1}
\end{equation}

\noindent Also the fluid model must satisfy the $Zeldovich\,\, condition$ i.e $p_r/\rho <1$ and $p_t/\rho <1$ everywhere inside the star which gives the lower bound of $B$ as

\begin{equation}
\left(\,\frac{a\,\cosh(b)}{c}-\tan^{-1}\sinh(b)\,\right)\,\,<\,\,B.   \label{B2}
\end{equation}

\noindent From Eq.(\ref{B1}) and Eq.(\ref{B2}), we get the following inequality

\begin{equation}
\left[\,\frac{a\,\cosh(b)}{c}-\tan^{-1}\sinh(b)\,\right]\,<\,\,B < \,\left[\,\frac{4\,a\,\cosh(b)}{c}-\tan^{-1}\sinh(b)\,\right].
\end{equation}

\begin{figure}[h]
\centering
\includegraphics[width=5.5cm]{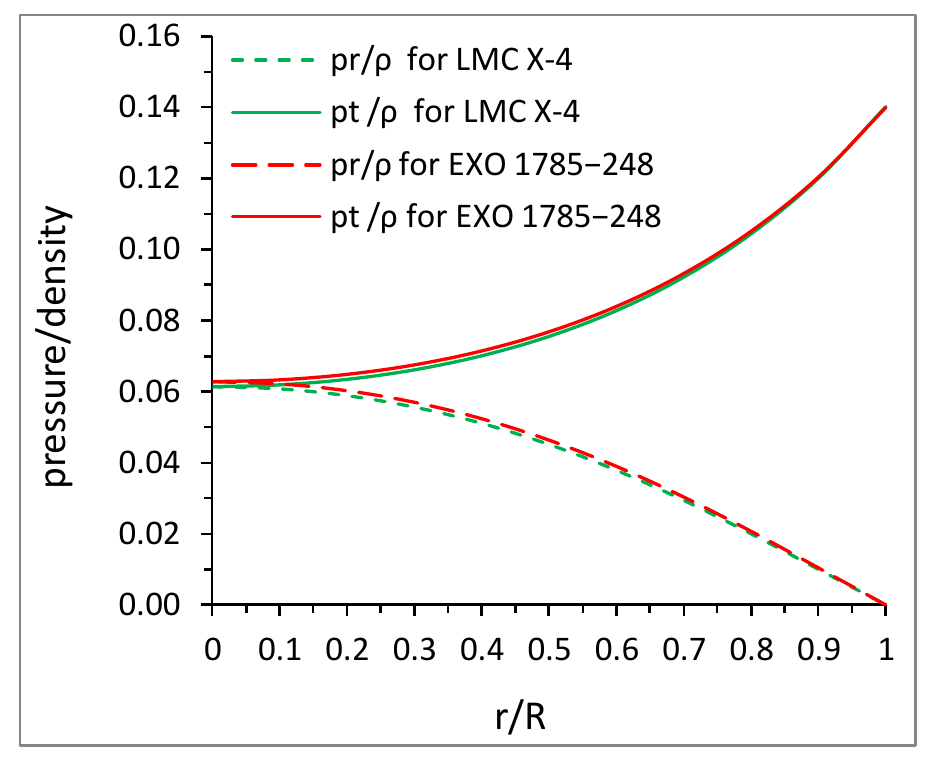}
\caption{Behavior of pressure-density ratio $p_i/\rho$ verses fractional radius $r/R$ for LMC X-4 and EXO 1785-248. For plotting of this figure the numerical values of physical parameters and constants are as follows:(i) $a=0.004$, $b=0.0021$, $c=0.0107$, $A=0.4806$, $B=1.2607$, $M=1.29 M_{\odot}$, and $R=8.831 km$ for LMC X-4, (ii). $ a=0.00393$, $b= 0.0025$, $c=0.01074$, $A=0.4905$, $B=1.2293$,  $M=1.3 M_{\odot}$ and $R=8.849 km$ for EXO 1785-248.}
\label{pd1}
\end{figure}

The behavior of $p_r/\rho$ and $p_t/\rho$ is shown in Fig.\ref{pd1}. From this figure it is clear that both $p_r/\rho$ and $p_t/\rho$ are less than 1 everywhere inside the anisotropic star which shows that our fluid model satisfies the $Zeldovich\,\, condition$.

\subsection{Matching conditions}
To find the arbitrary constants $A$ and $B$, we must match our interior solution to the exterior Schwarzschild solution at the boundary of the star. The line element of the exterior Schwarzschild solution is given by

\begin{equation}
	ds^{2} =\left(1-\frac{2M}{r} \right)\, dt^{2} -r^{2} (d\theta ^{2} +\sin ^{2} \theta \, d\phi ^{2} )-\left(1-\frac{2M}{r} \right)^{-1} dr^{2},
\end{equation}
where  the constant mass $M$ provides the total mass of the anisotropic star within $r=R$.
By matching of $e^{\nu}$ and $e^{\lambda}$ at the surface of the star ($r=R$) (continuity of first fundamental form) we get

\begin{equation}
	A\,\left[\,B + \tan^{-1}\sinh(a\,R^2+b)\,\right]^2=e^{\nu_R}=1-\frac{2M}{R} \label{nu1},
\end{equation}

\begin{equation}
	 \frac{1+\cosh[2(aR^2+b)]}{1+2cR^2+\cosh[2(aR^2+b)]} =e^{-\lambda_R}=1-\frac{2M}{R}. \label{nu2}
\end{equation}

\noindent For fixing the arbitrary constants, the second fundamental form ($\partial g_{44}/\partial r$) has to be also matched at the boundary $r=R$. By matching of $\partial g_{44}/\partial r$ at the surface of the star we get zero radial pressure at the boundary \cite{Misner1964}. Then $p_r=0$ at $r=R$ provides the value of arbitrary constant

\begin{equation}
	B=\frac{-c\,\tan^{-1}\sinh(aR^2+b)+4\,a\,\cosh(aR^2+b)}{c}. \label{b1}
\end{equation}

\noindent Then using Eqs.(\ref{nu1}) and (\ref{nu2}) together with Eq.(\ref{b1}), we obtain the values of constant $A$ and total mass $M$ as
\begin{equation}
	A=\frac{\cosh(a\,R^2+b)}{\sqrt{\cosh^{2}(aR^2+b)+c\,R^2\,}\,[\,B + \tan^{-1}\sinh(aR^2+b)\,]},
\end{equation}

\begin{equation}
	M=\frac{c\,R^3}{1+2c\,R^2+\cosh(2aR^2+2b)}.
\end{equation}

\begin{table}
\centering \caption{Numerical values of physical parameters
$a$, $b$, $c$, $R~(km)$, $M\left(M_\odot\right)$ for different
values of $n$~\cite{Gangopadhyay} \label{Table 1}}

{\begin{tabular}{@{}ccccccccc@{}}

Compact stars & $R~(km)$ &$M\left(M_\odot\right)$ &$a\,(km^{-2})$ & $b$ & $c\,(km^{-2})$ & $A$ & $B$    \\

\hline LMC X-4 & 8.831  & 1.29   & 0.004 & 0.0021&  0.0107 & 0.4806 & 1.2607 \\

\hline EXO 1785-248 & 8.849 & 1.3  & 0.00393 & 0.0025 & 0.01074 &  0.4905 & 1.2293 \\ \hline

\end{tabular}}
\end{table}

\begin{table}
	\centering
	\caption{The central density, surface density, central pressure and mass-radius ratio for compact star candidates}\label{Table 2}
	
\begin{tabular}{ccccc}
\hline
Compact star & Central density & Surface density & Central pressure &  $M/R$  \\
candidates & $(gm/cm^{3}) $ & $(gm/cm^{3})$ & $(dyne/cm^{2})$   \\ \hline

LMC X-4 & $ 1.7238\times{{10}^{15}}$ & $5.7038\times{{10}^{14}}$ & $9.526\times{{10}^{34}}$ & 0.2155 \\ \hline

EXO 1785-248 & $1.7302\times{{10}^{15}}$ & $5.7172\times{{10}^{14}}$ & $9.773\times{{10}^{34}}$ & 0.2167 \\ \hline

\end{tabular}
\end{table}

\section{Salient features of anisotropic models}

\subsection{Well behaved property of the solution}

For well behaved nature of the solution, the velocity of sound must be less than the velocity of light, and it should decrease monotonically throughout  the anisotropic star. For this purpose we have to calculate the radial and tangential speed of sound as

\begin{equation}
v_r=\sqrt{\frac{dp_r}{d\rho}}=\sqrt{\frac{dp_r/dr}{d\rho/dr}},
\end{equation}

\begin{equation}
v_t=\sqrt{\frac{dp_t}{d\rho}}=\sqrt{\frac{dp_t/dr}{d\rho/dr}}.
\end{equation}

\begin{figure}[h]
\centering
\includegraphics[width=5cm]{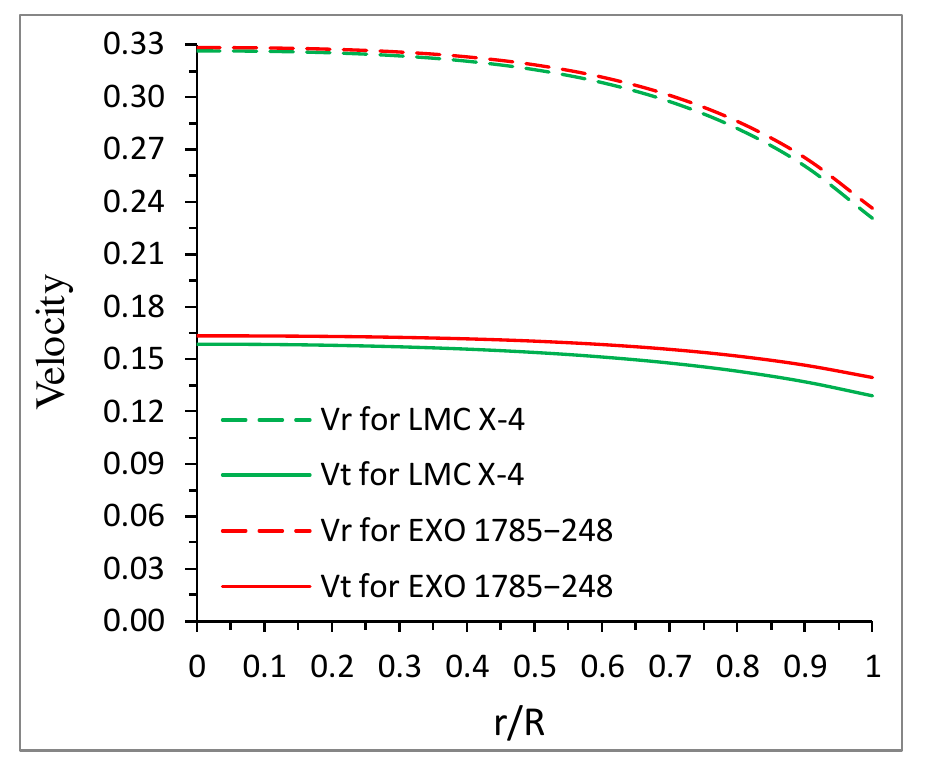}
\caption{Behavior of radial velocity, $v_r$, (dotted line) and tangential velocity, $v_t$, (solid line) verses fractional radius $r/R$ for LMC X-4 and EXO 1785-248. For plotting of this figure we have employed data set values of physical parameters and constants same as used in fig. 5.}
\end{figure}

\subsection{Dominant energy conditions}
For a physically reasonable anisotropic solution the energy momentum tensor has to obey the following dominant energy conditions:\\
(i) Null energy condition (NEC) implies that local mass-energy density must not be negative i.e. $\rho\geq 0$,\\
(ii) Weak dominant energy condition (WDEC) implies that the flow of energy inside star must not be faster than the velocity of light: $\rho-p_r \geq 0,~\rho-p_t\geq 0$,\\
(iii) Strong dominant energy condition (SDEC) implies that the flow of energy inside star must not faster than one-third of the light velocity: $\rho-3p_r\geq 0$, $\rho-3p_t\geq 0$.

\begin{figure}[h]
\centering
\includegraphics[width=5cm]{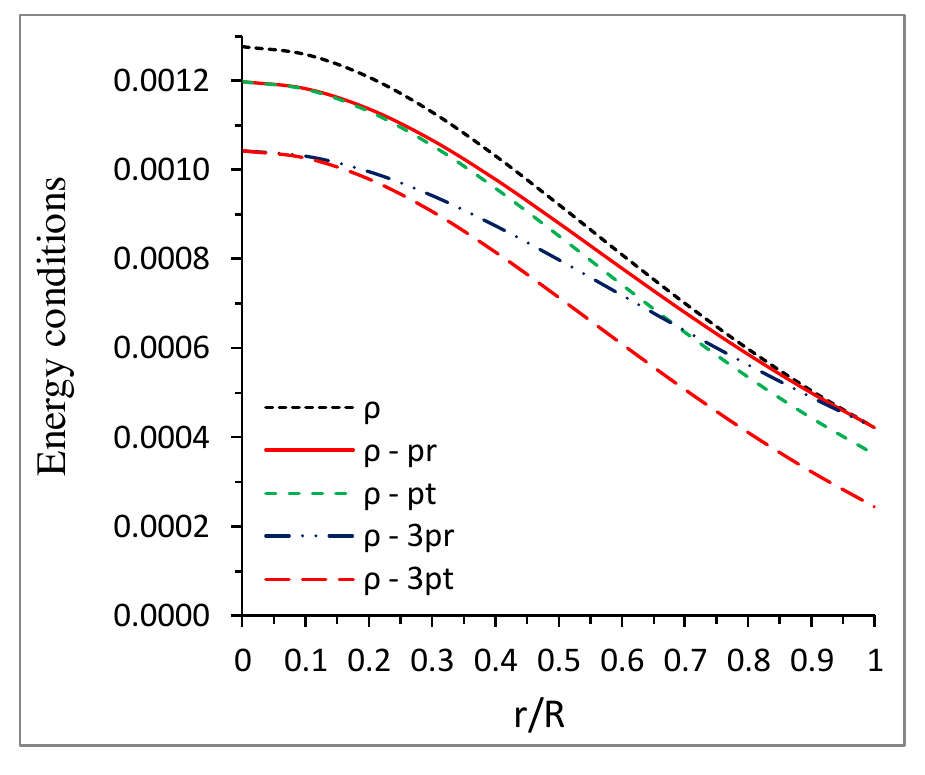} \includegraphics[width=5cm]{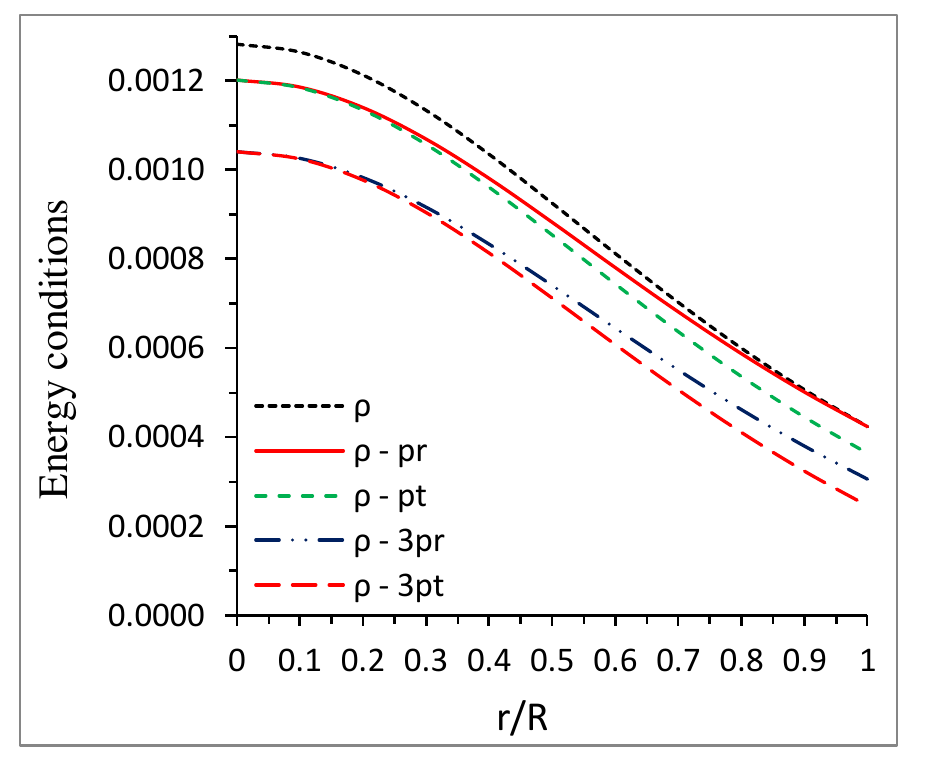}
\caption{Behavior of dominant energy conditions verses fractional radius $r/R$ for LMC X-4 (left panel) and EXO 1785-248 (right panel). For plotting of this figure the numerical values of physical parameters and constants are as follows:(i) $a=0.004$, $b=0.0021$, $c=0.0107$, $A=0.4806$, $B=1.2607$, $M=1.29 M_{\odot}$, and $R=8.831 km$ for LMC X-4, (ii). $ a=0.00393$, $b= 0.0025$, $c=0.01074$, $A=0.4905$, $B=1.2293$,  $M=1.3 M_{\odot}$ and $R=8.849 km$ for EXO 1785-248.}
\label{energy1}
\end{figure}

\noindent From Fig.\ref{energy1} it is clear that all energy conditions are satisfied within the compact star.

\subsection{Mass-radius relation}
For any physical valid star model according to Buchdahl \cite{Buchdahl} the mass to
radius ratio for perfect fluid model should be $2M/R> 8/9$. Later on Mak and Harko \cite{Mak2003} have proposed this relation in a
more generalized form which can be written as

\begin{equation}
M_{eff}=\frac{\kappa}{2}\int_0^{R}\rho\, r^{2}dr=\frac{c\,R^3}{1+2c\,R^2+\cosh(2aR^2+2b)}.  \label{eq30}
\end{equation}

 \noindent In this connection we would like to compare our proposed compact star model with the observed data of different realistic
objects. For this purpose we have calculated physical parameters for the models (see Tables \ref{Table 1} - \ref{Table 2}) by taking the mass of
the compact stars LMC X-4 and EXO 1785-248. The obtained radii of the different compact stars are given in Table \ref{Table 1} which are good agreement with the proposed values of Gangopadhyay et al.\cite{Gangopadhyay}.

\subsection{Surface redshift}

As we know the compactification parameter of the compact star is given by

\begin{equation}
u(R)=\frac{m(R)}{R}=\frac{c\,R^3}{1+2c\,R^2+\cosh(2aR^2+2b)}.\label{eq31}
\end{equation}

\begin{figure}[!htbp]\centering
    \includegraphics[width=5.5cm]{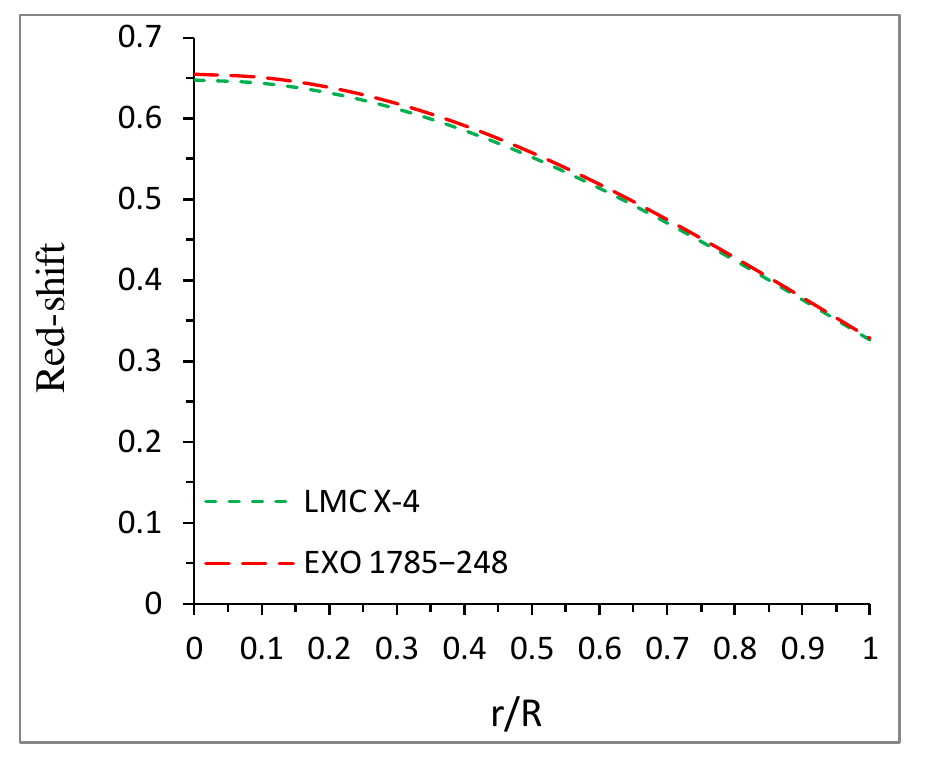}
\caption{Variation of redshift ($Z$) with the fractional coordinate $r/R$.For plotting of this figure the numerical values of physical parameters and constants are as follows:(i) $a=0.004$, $b=0.0021$, $c=0.0107$, $A=0.4806$, $B=1.2607$, $M=1.29 M_{\odot}$, and $R=8.831 km$ for LMC X-4, (ii). $ a=0.00393$, $b= 0.0025$, $c=0.01074$, $A=0.4905$, $B=1.2293$,  $M=1.3 M_{\odot}$ and $R=8.849 km$ for EXO 1785-248.}
    \label{Fig10}
\end{figure}

\noindent Then in connection to the above compactification parameter the surface redshift $(z_s)$ can be obtained directly by

\begin{equation}
z_s=\frac{1-[1-2u]^{\frac{1}{2}}}{[1-2u]^{\frac{1}{2}}}=\sqrt{\frac{1+2cR^2+\cosh[2(aR^2+b)]}{1+\cosh[2(aR^2+b)]}}-1. \label{eq32}
\end{equation}
From the above Eq.(\ref{eq32}) we can see that the surface redshift cannot be arbitrarily large as it depends upon the compactness parameter $u=m/R$. The behavior of the redshift inside the star can  see in  Fig. \ref{Fig10} which can be obtained by the formula $z=e^{-\nu/2}-1$.

\subsection{Stability of the solution}

\subsubsection{Stability of anisotropic models via cracking}
In our anisotropic fluid model, to verify stability we plot the radial
($v_{r}=\sqrt{dp_r/d\rho}$) and transverse ($v_{t}=\sqrt{dp_t/d\rho}$) sound speeds
in Fig. 5. It can be observed that both velocities satisfy the
inequalities $0< v^2_{r} < 1$ and $0< v^2_{t} < 1$ everywhere within the stellar object (Fig. 8) which is obeying the anisotropic fluid models \cite{Herrera1992,Abreu2007}.

\begin{figure}[h]
\centering
\includegraphics[width=5.5cm]{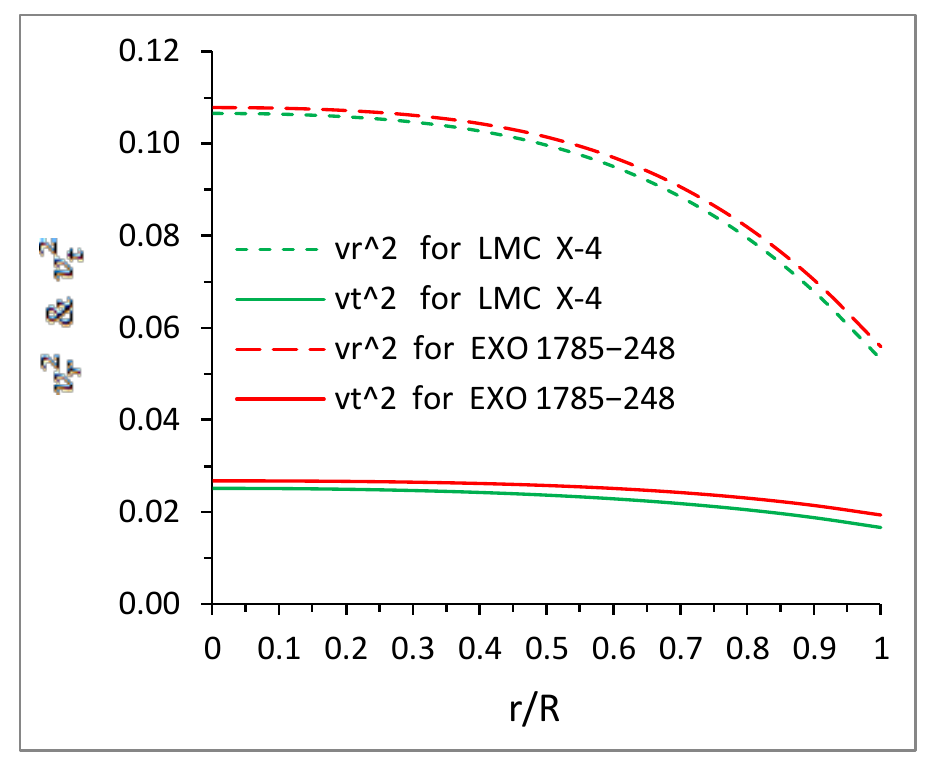}
\caption{Behavior of square of radial velocity, $v^2_r$,(dotted line) and tangential velocity, $v^2_t$,(solid line)  verses fractional radius $r/R$ for LMC X-4 and EXO 1785-248. For plotting of this figure the numerical values of physical parameters and constants are as follows:(i) $a=0.004$, $b=0.0021$, $c=0.0107$, $A=0.4806$, $B=1.2607$, $M=1.29 M_{\odot}$, and $R=8.831 km$ for LMC X-4, (ii). $ a=0.00393$, $b= 0.0025$, $c=0.01074$, $A=0.4905$, $B=1.2293$,  $M=1.3 M_{\odot}$ and $R=8.849 km$ for EXO 1785-248.}
\end{figure}

To check whether local anisotropic matter distribution is
stable or not, we use the proposal of Herrera \cite{Herrera1992},
known as cracking (or overturning) of the star. This indicates if the
region is potentially stable where the radial velocity of sound
is greater than the transverse velocity of sound. We can easily say that $0 < v^2_{r}-v^2_{t} < 1 $ (dotted line) and $-1 < v^2_{t} - v^2_{r} < 0$ (solid line) as can be seen from Fig. 9. Hence, we can conclude that our compact star model provides a stable configuration.

\begin{figure}[h]
\centering
\includegraphics[width=5.5cm]{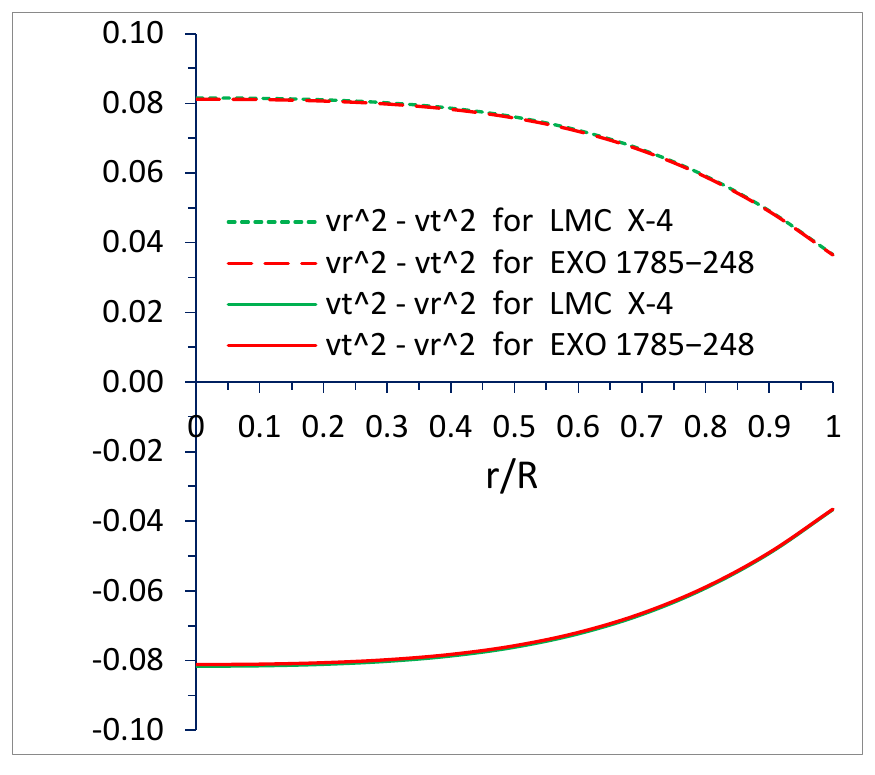}
\caption{Behavior of  $v^2_r-v^2_t$ (dotted line) and  $v^2_t-v^2_r$ (solid line) verses fractional radius $r/R$ for LMC X-4 and EXO 1785-248. For plotting of this figure we have employed data set values of physical parameters and constants which are the same as used in fig. 8.}
\end{figure}

\subsubsection{Stability via adiabatic index}
The stability of the relativistic anisotropic star also depends upon the adiabatic index $\Gamma$.
 Heintzmann and Hillebrandt \cite{Heintzmann1975} proposed that neutron
star models with anisotropic equation of state are stable if $\gamma >
4/3$.  However according to Newton's theory of gravitation the isotropic neutron star model
has no upper mass limit for the adiabatic index $\gamma > 4/3$ (\cite{bondi}). The adiabatic index $\Gamma$ is defined by

\begin{equation}
\Gamma=\frac{p_r+\rho}{p_r}\,\frac{dp_r}{d\rho}.
\end{equation}

\begin{figure}[h]
\centering
\includegraphics[width=5.5cm]{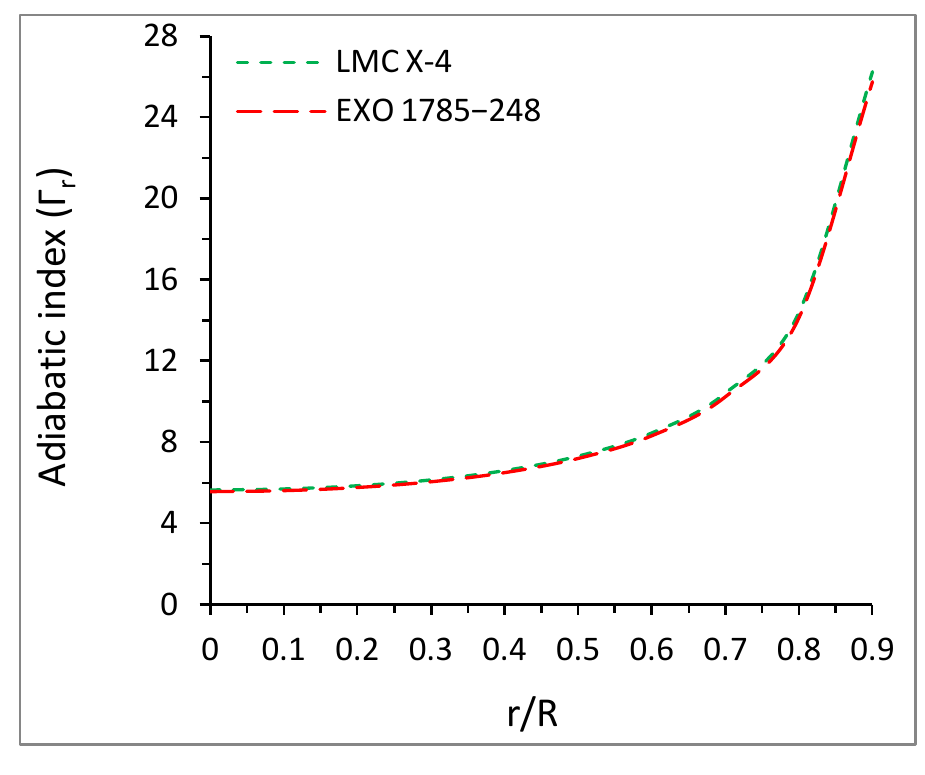}
\caption{Behavior of adiabatic index $\Gamma_r$ verses fractional radius $r/R$ for LMC X-4 and EXO 1785-248. For plotting of this figure we have employed data set values of physical parameters and constants which are the same as used in fig. 8 and 9.}
\label{gr}
\end{figure}

\noindent For a relativistic anisotropic fluid sphere the stability condition is given by
\begin{equation}
\Gamma>\frac{4}{3} \left[1 +3\pi\frac{\rho_0p_{r0}}{|p_{r0}^\prime|}r+\frac{(p_{t0}-p_{r0})}{|p_{r0}^\prime|r}\right],
\end{equation}
where, $p_{r0}$, $p_{t0}$, and $\rho_0$ denote the initial radial pressure, tangential pressure and energy density respectively in static equilibrium condition which satisfies the TOV equation (\ref{TOV}). However the second and last terms inside the square brackets correspond to the anisotropic and relativistic corrections (being positive quantities), which increase the unstable range of the adiabatic index. For this purpose we have plotted $\Gamma$ w.r.t. $r/R$. The behavior of the adiabatic index is shown in Fig.\ref{gr} which shows that  $\Gamma >\frac{4}{3}$ everywhere inside the compact star model.

\subsubsection{Stable equilibrium condition via TOV equation}
The Tolman-Oppenheimer-Volkov (TOV) equation describes the interior structure of the compact star which is a relationship between two physical quantities, the radial pressure and the density. Using the TOV equation we want to examine whether our present model is in a stable equilibrium stage under the three forces, viz. anisotropic force ($F_a=2(p_t-p_r)/r$), hydrostatics force ($F_h=-dp_r/dr$) and gravitational force ($F_g=-\nu'\,(\rho+p_r)/r$). This implies that the sum of three three different forces becomes zero:

\begin{equation}
F_a+F_h+F_g=0
\end{equation}

\noindent The explicit form of the above three different forces for this solution is given by

\begin{equation}
F_a=\frac{8\,r\,(c\,\Phi(r)-2\,a\,\cosh\psi)(c+a\,\sinh2\psi)}{8\,\pi\,(B+\tan^{-1}\sinh\psi)\,(1+2\,c\,r^2+\cosh2\psi)^2},
\end{equation}

\begin{equation}
F_h=-\frac{dp_r}{dr},
\end{equation}

\begin{equation}
F_g=-\frac{16\,a\,r\,\left[\,c\,\Phi(r)\,\cosh\psi+a\,p_2-2\,a\,c\,r^2\,\Phi(r)\,\sinh\psi \,\right]}
{8\,\pi\,(B+\tan^{-1}\sinh\psi)^2\,(1+2\,c\,r^2+\cosh2\psi)^2}.
\end{equation}

\begin{figure}[h]
\centering
\includegraphics[width=5cm]{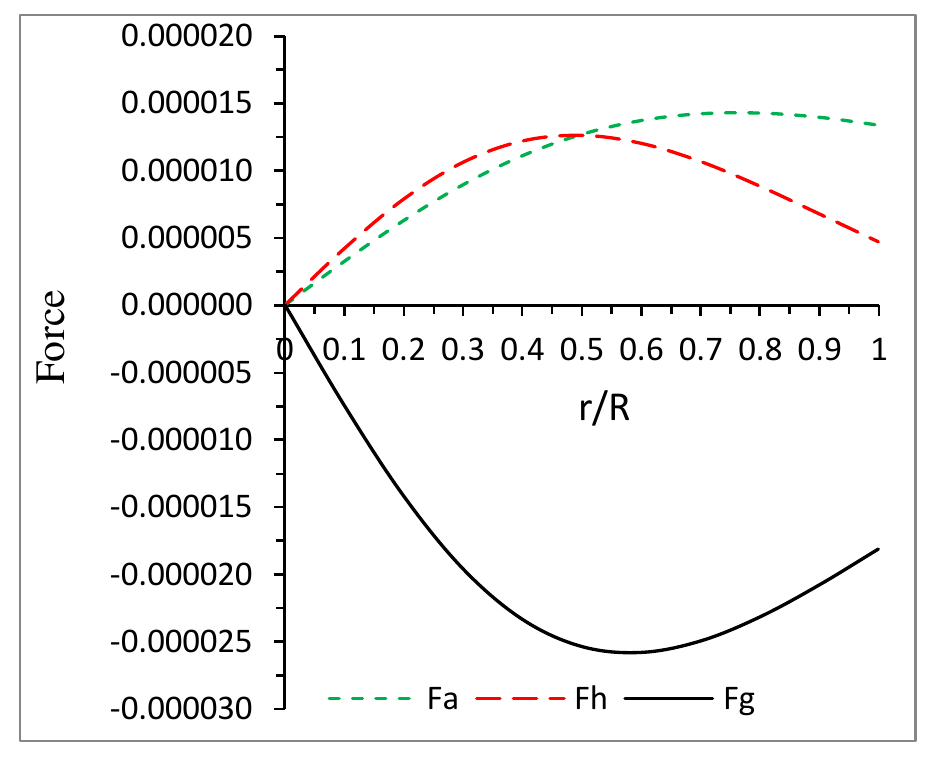} \includegraphics[width=5cm]{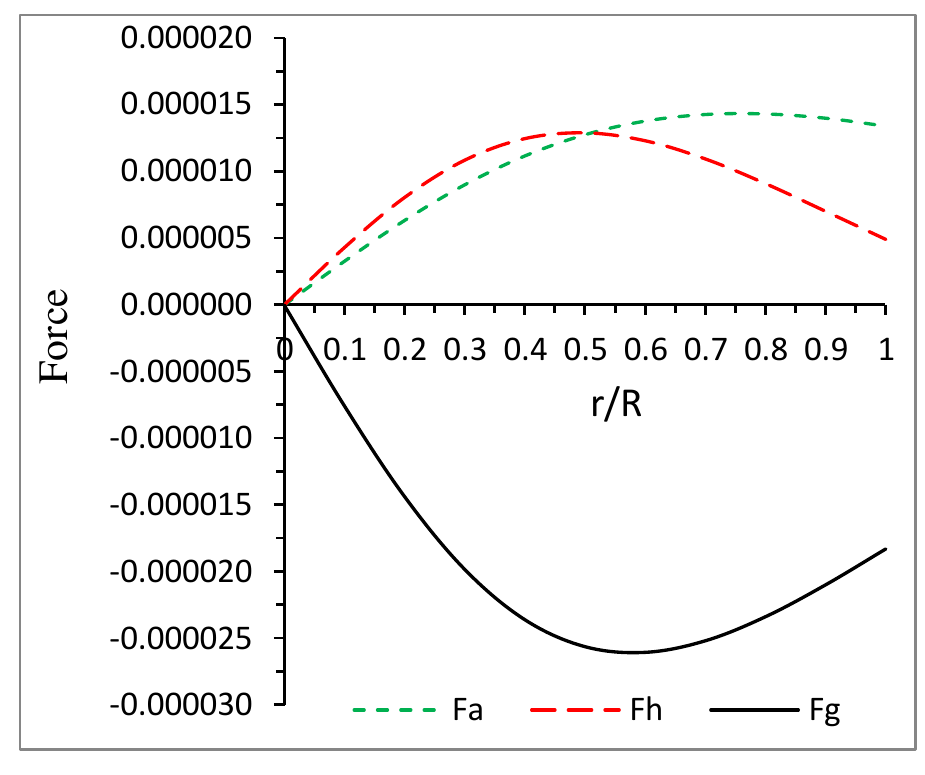}
\caption{Behavior of different forces verses fractional radius $r/R$ for LMC X-4 (left panel) and EXO 1785-248 (right panel). For plotting of this figure the numerical values of physical parameters and constants are as follows:(i) $a=0.004$, $b=0.0021$, $c=0.0107$, $A=0.4806$, $B=1.2607$, $M=1.29 M_{\odot}$, and $R=8.831 km$ for LMC X-4, (ii). $ a=0.00393$, $b= 0.0025$, $c=0.01074$, $A=0.4905$, $B=1.2293$,  $M=1.3 M_{\odot}$ and $R=8.849 km$ for EXO 1785-248.}
\end{figure}

\section{Physical analysis and discussion}
In the present paper we have investigated the nature of anisotropic fluid spheres by utilizing
the Karmarkar condition which are also known as  spacetimes of embedding class one. To outline this investigation
we have considered the following assumption for the gravitational potential:
 $\lambda =  \ln[1+2\,cr^2+\cosh(2\,ar^2+2\,b)]-\ln[1+\cosh(2\,ar^2+2\,b)]$,
 where $a$ and $c$ are nonzero positive parameters. The choices of $a$ and $c$ are as follows:
 (i). If $c=0$ then we have flat spacetime.
 (ii). If $a=0$ then there is no pressure-free boundary unless $c=0$.
Under the above restrictions, we have obtained new anisotropic fluid spheres for $a\ne0$ and $c\ne0$.

The main physical features of the present solution can be used to explore the nature of anisotropic fluid spheres as follows:

(i) Firstly we have developed a relation between the gravitational potentials $e^{\nu}$ and $e^{\lambda}$ for the spherically symmetric line element by using the Karmarkar condition. Due to this relation, we can convert all the differential equations in terms of one of the metric coefficients (the full details can see form the reference Maurya et al. \cite{maurya11,maurya12}). For this purpose we have assumed totally new metric potential $e^{\lambda}=[{1+2c\,r^2+\cosh2(ar^2+b)}]/[{1+\cosh2(ar^2+b)}]$ to find the anisotropic solution for realistic fluid spheres. The variation of $e^{\nu}$ and $e^{\lambda}$ can be see from Fig.1.

(ii). The fluid spheres are purely anisotropic because embedding class one solutions can give only two types of prefect fluid solutions which are namely the Kohlar-Chao or Schwarzschild otherwise the metric turns out to be flat. As we can see the radial pressure $p_r$ is zero at boundary but the tangential pressure $p_t$ does not vanish at $r=R$; however both are equal at the centre of the fluid sphere (Fig.2). Also the pressure anisotropy vanishes everywhere inside the fluid sphere if and only if $c=0$. In this situation the pressures and density become zero and the metric turns out to be flat. As we can see form Fig.3 the anisotropy is increasing throughout while it is zero at the centre, which implies that $p_r$ and $p_t$ are not equal except at the centre of the fluid sphere.

(iv). Since $p_r/\rho$ and $p_t/\rho$ lie between 0 and 1 everywhere within the sphere, our fluid sphere satisfies the $Zeldovich\,\, condition$. Moreover $p_r/\rho$ and $p_t/\rho$ are monotonically decreasing throughout the inside the sphere (Fig. 4).

(v) For the well behaved nature of the solution the velocity of sound should be decreasing throughout the fluid sphere and must be less than the velocity of light. From Fig.5, it is clear that both $v_r$ and $v_t$ are decreasing and less than 1 which shows that our anisotropic solution is well behaved. Also all dominant energy conditions are satisfied necessary physical requirement everywhere inside the fluid sphere (Fig.6).

(vi) The surface redshift is also determined by using the compactness factor for the fluid sphere. For the fluid sphere EXO 1785-248, the surface redshift turns out to be $z_s=0.3285$ which is a maximum. The redshift without cosmological constant for isotropic fluid spheres is bound by $z_s\le 2$ \cite{Buchdahl,Straumann,Böhmer}. However in the presence of the cosmological constant Bohmer and Harko \cite{Böhmer} argued that the surface redshift must satisfy the restriction $z_s\le 5$ for anisotropic fluid spheres. Therefore the value of surface redshift for our anisotropic spheres seems to be compatible with realistic compact objects. Moreover the surface redshift cannot be arbitrarily large because it depends on the compactness factor $u=M/R$.

 (vii) We have also discussed the stability of the fluid sphere (which is the most vital physical requirement) by using the following facts: (a) the cracking concept proposed by Herrera \cite{Herrera1992}, (b) the variation of adiabatic index inside the fluid spheres, (c) the stable equilibrium condition by using TOV equation. The results are as follows: It can be observed from Fig.(8) that the velocity $v^2_r$ and $v^2_t$ are lying in the range $0.053\le v^2_r\le0.107$, $0.017\le v^2_t\le 0.025$ for LMC X-4 and $0.056\le v^2_r\le0.11$, $0.019\le v^2_t\le 0.027$ for EXO 1785-248. Also the radial velocity ($v_r$) is always greater than the tangential velocity ($v_t$) everywhere inside the fluid sphere (Fig.5). We plot the figure for $v^2_r-v^2_t$ and $v^2_t-v^2_r$ to apply the Harrera cracking concept observing that there is no change in sign of $v^2_r-v^2_t$ and $v^2_t-v^2_r$ (Fig.9). This implies that our anisotropic fluid models are stable. The variation of adiabatic index is given in Fig.10 which shows that the value of $\Gamma$ is more than $4/3$ within the fluid models. Finally we discuss the equilibrium condition  for the anisotropic fluid model by using the Tolman-Oppenheimer-Volkoff (TOV) equation. For this purpose we plot figures for TOV equation in terms of different forces. From Fig.11, we can observe the gravitational force $F_g$ is counter-balance by the joint action of hydrostatic force $F_g$ and the anisotropic force $F_a$ for both LMC X-4 and EXO 1785-248.

In this paper we have shown that the Karmarkar embedding condition  describes a rich class of anisotropic compact spheres which are physically viable in relativistic astrophysics.

\textbf{Acknowledgments}: The author S. K. Maurya et al. acknowledges continuous support and encouragement from the administration of University of  Nizwa. SDM acknowledges that this work is based upon research supported by the south African Research Chair Initiative of the Department of Science and Technology and the National Research Foundation.

\end{document}